\documentstyle[preprint,aps]{revtex}
\begin{document}

 \baselineskip 19pt


\title {ORIGIN OF THE HIGH ENERGY
EXTRAGALACTIC DIFFUSE GAMMA RAY BACKGROUND}
\author{Arnon Dar
 \footnote{e-mail: PHR19AD@technion.technion.ac.il}
 and Nir J. Shaviv
 \footnote{e-mail: shaviv@tx.technion.ac.il}
}

\address{Department of Physics and Space Research Institute
Technion - Israel Institute of Technology, Haifa 32000, Israel}
\maketitle
\begin{abstract}
We show that cosmic rays in external galaxies, groups and
clusters rich in gas, with an average flux similar to that observed
in the Milky Way, could have produced the observed extragalactic diffuse
gamma radiation.

 \vskip 1.0 cm
 \centerline{\it Submitted to Physical Review Letters}

\end{abstract}


In addition to the galactic diffuse gamma radiation, which varies
strongly
with direction and can be explained by cosmic ray interactions in the
galactic interstellar medium\cite{BL}
, there appears
to be an unaccounted diffuse component which is isotropic at least on a
coarse scale and fits well at low photon energies to the extragalactic
hard X-ray background radiation\cite{EC}.
These two features suggest an extragalactic origin \cite{EC} of the isotropic
diffuse gamma ray background radiation (GBR). Various unresolved
extragalactic discrete and diffuse sources of gamma rays have been
suggested\cite{RM} to explain its origin, its observed flux
level (($5.5\pm 1.3)\times 10^{-5}$ and $(1.3\pm 1.3)\times 10^{-5}~
\gamma~cm^{-2}~s^{-1}~ster^{-1}$ above 35 MeV and 100 MeV,
respectively) and its observed \cite{DJ} spectral features (a power-law
spectrum with a power index
 $\alpha=2.4^{+0.4}_{-0.3}$
between a few tens keV and a few hundreds Mev with a possible bump
around a few MeV). The discrete sources included active galactic
nuclei (AGN), Seyfert galaxies, BL Lac objects and normal galaxies
while the diffuse sources included inverse Compton scattering of the
microwave background photons in intergalactic space from cosmic ray
electrons and more speculative sources  such as matter-antimatter
annihilation, cosmic strings, annihilation of supersymmetric particles
and decay of particles relics from the Big-Bang. However, all these
explanations have been later questioned by observations.
For instance, one of the more promising explanations of the GBR has been
that it is the sum of gamma ray emission from unresolved AGN\cite{AW}.
When this explanation was suggested, only the relatively nearby
quasar 3C 273 had been seen in high energy gamma rays\cite{BN}
$(E>100~MeV)$. Since the launch
of the Compton Gamma Ray Observatory (CGRO) in 1991, its Energetic
Gamma Ray Experiment Telescope (EGRET) has detected 33 bright AGN
in high energy gamma rays\cite{VM},
all of which seem to belong to the ``blazar'' class. Many more
AGN including blazars, which are both bright and relatively close
and which were within the EGRET field of view, have not been detected
in high energy gamma rays, indicating that not all of these objects
are so luminous in gamma rays, or their emission is highly beamed or
has a low duty cycle. The beaming hypothesis is further supported by
other features of the sources such as superluminal velocity, beamed radio
emission, and a power-law gamma ray spectrum (with a power index between
1.4  and 3.0, with values between 1.8 and 2 being most common). It is
therefore difficult to see how the observed isotropic GBR can be produced
by the highly beamed emission from AGN, many of which have
a much  harder spectrum than that of the GBR\cite{CD}.
\smallskip

Most of the high energy gamma ray emission of our Milky Way (MW) galaxy
can be explained by cosmic ray interactions with its interstellar
gas \cite{BL}. While the total mass of gas in the MW is only about 10\% of
its stellar mass, i.e., $M_{gas}/M_*\approx 0.10$, recent
X-ray observations of clusters and groups indicate that they contain
much larger fractions of gas\cite{LP}.
For instance, analyses of recent observations
with the ROSAT X-ray telescope of the compact group HCG62 and the Coma
cluster yielded $M_{gas}/M_*\approx 1.1h^{-3/2}$ within a distance of
$0.24h^{-1}~Mpc$ from the center of HCG62\cite{TJ},
and $M_{gas}/M_*\approx (9.0
\pm 2.6)h^{-3/2}$ within a distance of $2.5h^{-1}~Mpc$ from the center
of the Coma cluster\cite{UGB},
where $H_0=100h~km~s^{-1}Mpc^{-1}$ is the Hubble constant.
Detailed studies of the gas content of galaxies, groups and clusters
indicate that for the whole Universe $M_{gas}\sim M_*$, i.e.,
$M_{gas}/M_*=\beta(M_{gas}/M_*)_{MW}$ with $\beta\sim 10$.
At higher redshifts a larger fraction of the baryons
in the Universe are expected to reside in gas because cosmic
evolution transforms gas into galaxies and
stars. The gas content of the Universe at high redshift has
been determined from the statistics of Lyman $\alpha$ absorption
systems in quasar spectra measured with the large ground based
telescopes and the Hubble Space Telescope. It is consistent with
the present amount of visible matter in galaxies and clusters.
In this letter we show that if the flux of high energy cosmic ray
nuclei is either universal and equal to that observed in the MW,
or if the MW cosmic ray flux represents well the average cosmic
ray flux in clusters and groups of galaxies, then the observed
extragalactic GBR could have been produced by cosmic ray interactions
with gas in external galaxies, groups and
clusters\cite{SMC}.
\smallskip

We first present two simple
estimates of the extragalactic GBR produced by cosmic ray
interactions in extragalactic gas which ignore  cosmic evolution.
Cosmic evolution effects (e.g., Hubble expansion, energy redshift,
gas consumption by stellar formation, evolution of the cosmic ray flux)
generally tend to increase the predicted GBR and
will be included later.
\smallskip

The main mechanisms by which cosmic rays produce high energy gamma -
rays in the interstellar medium are bremsstrhalung and inverse Compton
scattering from electrons and radiative decay of
$\pi^0$'s produced in inelastic collisions of cosmic ray nuclei
(mainly protons) with the interstellar gas. Bremsstrahlung dominates
gamma ray production at low energies while $\pi^0$ production
dominates production of high energy gamma rays. Thus, for a universal
cosmic ray flux, the emission is proportional to the mass of
the gas (assuming a universal composition). The observed \cite{BL} total
luminosity of the galaxy in gamma rays with energy above 100 MeV
is $L_\gamma=(2\pm 1)\times 10^{39}~erg~s^{-1}.$ The optical
luminosity of the galaxy was estimated \cite{VDB} to be
 $(2.3\pm 0.6)\times
10^{10}L_\odot$,
while the average luminosity density of the Universe was estimated \cite{JL} to
be
$(1.83\pm 0.35)\times 10^8h^2L_\odot~Mpc^{-3}$.
Consequently,
cosmic ray interactions in extragalactic gas produce a GBR with
an energy spectrum similar to that of the galactic diffuse gamma
radiation and a flux level of,
\begin{equation} F_\gamma(>E)\approx  {\rho_L\over L_{MW}}
                 { c\beta L_\gamma(>E)\over 4\pi H_0}.
\end{equation}

It yields, e.g.,
$F_\gamma(>100~MeV)\approx (5.0\pm 3.0) \times 10^{-9}~erg~cm^{-2}s^{-1}
ster^{-1}, $  where we used the value,
$h\approx 0.80\pm 0.16$, that was measured recently
with the Hubble Space Telescope\cite{RFR}.
This flux level agrees, within
the experimental uncertainties, with the flux level of the observed GBR.
\smallskip

The large formal uncertainty in this estimate is due to the
uncertain values of the total gamma ray and optical luminosities of
the Milky Way and the value of the Hubble constant.
If most of the baryons in the present
Universe are still in the gas state then
these uncertainties can be avoided by using an alternative estimate:
\begin{equation}
F_\gamma (>E)\approx cH_{0}^{-1}n_bj_{_H}(>E).
\end{equation}
For $j_{_H}(E>100~MeV)=(2.6\pm 0.24)\times
                              10^{-26}~\gamma~cm^{-2}s^{-1}ster^{-1}$
(see ref. \cite{BL}),
where $n_b$ is the mean baryon density in the Universe and $j_{_H}(>E)$
is the total emissivity in high energy gamma rays per baryon
due to cosmic ray interactions.
$j_{_H}(>E)$ is
known quite accurately, both theoretically and experimentally \cite{BL},
from  cosmic ray interactions in the solar neighbourhood,
  $j_{_H}(E>100~MeV)=(1.8\pm 0.2)\times
            10^{-26}~\gamma~cm^{-2}s^{-1}ster^{-1},$  while
$n_b$ is known quite accurately from Big Bang nucleosynthesis (SBBN):
The ratio of baryons to photons in the Universe, $\eta=n_b/n_\gamma=
(1.6\pm 0.1)\times 10^{-10},$
was deduced \cite{RAD}
 from the observed
abundances of the light elements, H, D, $^4$He and $^7$Li extrapolated
to zero age using Standard Big-Bang Nucleosynthesis (SBBN) theory.
The microwave background radiation measured with the Cosmic
Background Explorer (COBE) satellite is well fitted with a black body
radiation of temperature \cite{RJM}
  $T=2.726\pm 0.017 ~K$, yielding
$n_\gamma=411\pm 8~cm^{-3}$. Consequently, SBBN yields
$n_b=(6.6\pm 0.4)\times 10^{-8}cm^{-3}$ and from
Eq.2 it follows that
\begin{equation}
F_\gamma(>100~MeV)\approx (1.4\pm 0.3)\times
                 10^{-5}~\gamma~cm^{-2}s^{-1}ster^{-1},
\end{equation}
which agrees with the flux level of the observed GBR.

Note that we used a mean baryon density which is implied \cite{SMC}
by the recent measurements of the abundance of Deuterium in
a high redshift low metallicity cloud\cite{RAS}
($(1.9-2.5)\times 10^{-4}$ by number)
and also by the current best determinations\cite{RJP}
of the primordial abundance of $^4$He ($22.9\pm 0.5\%$ by mass).
However, if the primordial abundance of Deuterium is that observed in
the local interstellar medium\cite{RJL}
 ($(1.65^{+0.07}_{-0.18})\times 10^{-5}$)
and if the current best determinations of the primordial abundances of
$^4$He and $^7$Li underestimate their true values
due to large systematic errors,
then $n_b\approx 2.5\times 10^{-7}~cm^{-3}$ and most of the baryonic matter in
the Universe is optically dark. Some of it, however, may reside in
low density (X-ray dim) diffuse ionized gas. In that case
Eq. 3 should be regarded as an estimated lower bound on the GBR.
\smallskip

We will now consider cosmic evolution effects. In a
Friedman-Robertson-Walker universe
with a zero cosmological constant and a density $\Omega$ in
critical density units (i.e.,
$\Omega=\rho/\rho_c$ where $\rho_c=3H_0^2/8\pi G$ with
G being Newton's constant of gravity),
a population of gamma ray sources $s$ with differential luminosities
$L_s(E)\propto E^{-\alpha}$, a number density at the present epoch
$n_s$ and evolution functions $f_s(z)$ (which describe how
the luminosities of sources $s$ change with redshift $z$),
generates a diffuse extragalactic flux
at the present cosmic epoch, $z=0$, which is given by
\begin{equation}
{dF\over dE}=\Sigma_s{L_s n_s \over 4\pi}{c\over H_0}\int_0^\infty
{f_s(z)dz\over (\Omega z+1)^{1/2}(1+z)^{2+\alpha}}~.
\end{equation}
In Eqs.1-4 we have neglected gamma ray absorption in the intergalactic
space which  becomes important only for $E_\gamma>500$ GeV due to
$e^++e^-$ pair production in collision with photons of the
intergalactic IR background radiation.
Besides the gas content of the Universe
the major unknown cosmological factor which affects Eq.4
is the evolution of the cosmic
ray flux. We have investigated various plausible cosmic evolutions
and found out that they can change the predicted
flux level of the GBR as given by Eq.2
by less than a factor two. For instance,
if high energy cosmic rays are universal and were injected into
the intergalactic space during a short time period around redshift $z_i$
$(1+z_i\gg 1)$, then their density at redshift $z<z_i$
was $(1+z)^3$ times larger. Consequently, the emissivity
of atoms was $(1+z)^3$ times larger and the cosmic integration
increases the  GBR by a factor
$K\approx(1+\Omega/2)/(\alpha+\Omega/2-2)$. The
cosmic ray differential flux below 1000 GeV
has  a power index of $\alpha\approx 2.67$, yielding
K= 1.28 for $\Omega=1$ and K= 1.49 for $\Omega=0$.
\smallskip

Fig.1 presents a comparison between
our prediction of the cosmic ray produced extragalactic
GBR for the choice $h=0.8$ and $\Omega=0.15$,
and the observed extragalactic GBR.
As explained, we estimate that the uncertainty
in the absolute normalization of the flux level of the GBR
due to uncertain cosmic evolution is less than a factor two.
As demonstrated in Fig.1, an average  cosmic ray flux in external,
gas-rich, galaxies, groups and clusters, similar to that observed
in the Milky Way can produce the observed extragalactic GBR.
\smallskip

Our explanation of the GBR can be further tested in future observations,
at least in two ways:
\smallskip

\noindent
a) At GeV energies gamma ray production  is dominated by $\pi^0$
production. Because of Feynman scaling of pion production at GeV
energies, the GBR should exhibit the same power-law spectrum exhibited
by the MW cosmic ray flux,
namely, $\alpha\sim 2.67$. Above 500 GeV the power index should
increase with increasing energy due to pair production off the
extragalactic IR background photons.

\noindent
b) Nearby gas-rich clusters which were observed
with the ROSAT X-ray telescope, such as the Coma cluster \cite{UGB}
($D\approx 70h^{-1}Mpc$)
and the Perseus cluster\cite{RA}
 ($D\approx 55h^{-1}Mpc$)
may be detected by EGRET in high energy
gamma rays. The expected gamma ray flux from the interaction of
a MW like cosmic ray flux in a gas rich cluster
at a distance $D$ from Earth is given by,
\begin{equation}
F_\gamma(>E)\approx {M_{gas}j_{_H}(>E)\over m_pD^2}.
\end{equation}
For the Coma cluster,
$M_{gas}=(9.0\pm 2.6) \times 10^{13}h^{-5/2}M_\odot$
within  a radius of 100 arcmin from its center \cite{UGB}.
Consequently, its expected gamma ray flux level at Earth is, $F_\gamma(
>100~MeV)\approx 0.5\times 10^{-7}~\gamma~cm^{-2}~s^{-1}. $
For the Perseus cluster,
$M_{gas}=(1.2\pm 0.3) \times 10^{14}h^{-5/2}M_\odot$
within  a radius of 100 arcmin from its center\cite{RCJ}.
Consequently, its expected gamma ray flux level at Earth is, $F_\gamma
(>100~MeV)\approx 1.0\times 10^{-7}h^{-1/2}~\gamma~cm^{-2}~s^{-1}. $
Such fluxes are marginally detectable by EGRET. However,
a positive detection by EGRET of time independent
gamma ray fluxes of these levels
from the directions of the Coma and Perseus clusters
will provide supportive evidence for the cosmic ray origin of the GBR.
Conversely, a failure to detect gamma rays from the Coma and Perseus
clusters by EGRET can be used to set a limit on the cosmic ray flux in
these clusters, on the validity of cosmic ray universality and on
the validity of the cosmic ray origin of the extragalactic GBR.

\begin{figure}
\noindent{ FIG. 1. Comparison between the predicted (dotted line)
extragalactic GBR produced by a universal MW-like cosmic ray flux
in extragalactic gas (Eq.2) and the observed \cite{EC,RM} high energy GBR
(dashed strip).}
\end{figure}
\end{document}